# Computer Simulation of the Early Stages of Self-Assembly and Thermal Decomposition of ZIF-8

S. R. G. Balestra[1, 2, 3] and R. Semino[*1, 4]
[1)]*ICGM, Univ. Montpellier, CNRS, ENSCM, Montpellier, France.*
[2)]*Departamento de Sistemas Físicos, Químicos y Naturales, Universidad Pablo de Olavide, Ctra. Utrera km 1, Seville ES-41013, Spain.*
[3)]*Instituto de Ciencia de Materiales de Madrid, Consejo Superior de Investigaciones Científicas (ICMM-CSIC) c/ Sor Juana Inés de la Cruz 3, Madrid ES–28049, Spain.*
[4)]*Sorbonne Université, CNRS, Physico-chimie des Electrolytes et Nanosystèmes Interfaciaux, PHENIX, F-75005 Paris, France*

(*Electronic mail: rocio.semino@sorbonne-universite.fr)

(Dated: October 19, 2022)

We employ all-atom well-tempered metadynamics simulations to study the mechanistic details of both the early stages of nucleation and crystal decomposition for the benchmark metal-organic framework ZIF-8. To do so, we developed and validated a force field that reliably models the modes of coordination bonds *via* a Morse potential functional form and employs cationic and anionic dummy atoms to capture coordination symmetry. We also explored a set of physically relevant collective variables and carefully selected an appropriate subset for our problem at hand. After a rapid increase of the Zn-N connectivity, we observe the evaporation of small clusters in favor of a few large clusters, that lead to the formation of an amorphous highly-connected aggregate. $Zn(MIm)_4^{2-}$ and $Zn(MIm)_3^-$ complexes are observed, with lifetimes in the order of a few picoseconds, while larger structures, such as 4-,5- and 6-membered rings, have substantially longer lifetimes of a few nanoseconds. The free ligands act as "templating agents" for the formation of the sodalite cages. ZIF-8 crystal decomposition results in the formation of a vitreous phase. Our findings contribute to a fundamental understanding of MOF's synthesis that paves the way to controlling synthesis products. Furthermore, our developed force field and methodology can be applied to model solution processes that require coordination bond reactivity for other ZIFs besides ZIF-8.

## I. INTRODUCTION

Metal-Organic Frameworks (MOFs) have revolutionized research in the fields of clean energy, environment and health due to their wide variety both in terms of chemical composition and topology, which confer them a vast array of shape–, selectivity–, reactivity–, and confinement–based properties. Indeed, MOFs can be formed by most of the existing metals in the periodic table in the form of cations or metal-based clusters combined with a large variety of organic polydentate ligands such as carboxylates, phosphonates and imidazolates, and they can exhibit many different pore shapes and sizes.[1] Thus, they constitute the ideal playground for synthesis scientists to create new porous materials with societally relevant applications, including water and food treatment, carbon capture, catalysis and drug delivery.[2–8] Even though researchers have both achieved a good understanding of structure/property relationships for MOFs[9–14] and mastered several MOF synthesis techniques,[15] MOF rational design is still in its infancy. Only a few valuable efforts have led to establishing some principles for the rational design of MOF synthesis based on coordination and reticular chemistry.[16–20] A deeper understanding of the mechanisms underlying MOF self-assembly would undoubtedly make a leap in the quest for MOF rational design.

While a few experimental studies have addressed the crystal growth of MOFs by various *ex situ* and *in situ* analytical techniques, including HRTEM, EDXRD and diverse spectroscopies,[21–30] the required resolution to fully understand this complex process at the molecular-level cannot yet be experimentally achieved, that is, the association of the ligands with the cationic moieties to form oligomeric aggregates in solution leading to the formation of the first secondary building units (SBUs) and beyond. From the modeling standpoint, only a handful of studies have been devoted so far to the simulation of the pre-nucleation process of MOFs in solution. Yoneya, Tsuzuki, and Aoyagi[31] modeled the first stages of polymerization between $Ru^{2+}$ and $Pd^{2+}$ and 4,4'-bipyridine by means of classical molecular dynamics (MD) simulations considering an implicit solvent model. Biswal and Kusalik[32] then applied the same methodology to simulate the polymerization between $Zn^{2+}$ and 1,4-benzenedicarboxylic acid ligands including an explicit solvent model, and further studied the effect of the complexity of the chosen model on the results.[33] Despite the fact that the importance of the force field was clearly demonstrated in this latter study, no special efforts were dedicated to the development and validation of force fields to model MOF self-assembly up to date. These studies have introduced the use of cationic dummy atoms (CDA) models,[34] in order to better represent the anisotropic electronic density distribution when the metal cation is exposed to a field generated by the organic ligands to reproduce the correct topology of the final material. Since then, CDA models have been extended to new MOF families.[35]

The above mentioned studies described the formation of amorphous aggregates comprising the inorganic and organic moieties in solution, but no signs of the onset of crystallization were observed. This is not surprising, since the collective motions that are required to achieve local order constitute rare events, which are only very seldom sampled in a regular MD simulation. It is only by the use of enhanced sampling tech-niques that these energy barriers may be overcome in a statistically meaningful fashion. Colón et al.[36] were able to obtain a cluster–like ordered structure by means of Finite Temperature String methods. Following the same philosophy, Kollias et al.[37] applied metadynamics[38] simulations to the study of the early stages of the nucleation of MIL-101(Cr) starting from half SBUs, and made a step forward by also considering the ionic force effects by means of the explicit inclusion of counterions. This latter work was a follow up of a previous ab initio investigation of the mechanism of formation of MIL-101(Cr) SBUs.[39] The same authors have very recently published a continuation of their MIL-101 work, where they develop a graph-based methodology to understand the influence of solvent and ions in the formation of the building units.[40] Also recently, Filez et al.[30] combined density functional theory (DFT) calculations and MD simulations with in situ elastic scattering and mass spectroscopy to unveil molecular-level details of the mechanism of the first stages of the nucleation of ZIF-67, a $Co^{2+}$ sodalite MOF. They describe the metal-ligand complexes that are formed in solution upon addition of the ligand as well as the associated Oh $\rightarrow$ Td symmetry changes and their lifetimes and give additional evidence to the hypothesis of the existence of an amorphous intermediate species along the self-assembly process.

The mechanism of MOFs' thermal decomposition, in turn, can be studied via direct experimental measurements such as variable temperature X-ray diffraction and differential scanning calorimetry[41] but to the best of our knowledge it has not yet been probed by molecular simulation techniques. Simulation can offer molecular level detail into the decomposition mechanisms for the first time, which could help guide their implementation as a defect engineering methodology.[42]

In this contribution, we study both (a) the early stages of self-assembly of metal ions and organic ligands and (b) the decomposition of a pre-formed MOF crystal, by Well–Tempered Metadynamics (WT-MetaD) simulations.[43,44] We introduce several methodological improvements to previous works: (i) a more physically sound force field that treats the metal-ligand interaction with a Morse potential instead of a Lennard-Jones or a harmonic potential as was the case in previous works, (ii) a robust force field validation, and (iii) a comprehensive exploration of the suitability of different collective variables for the study of nucleation.[45–47] We apply our simulation method to the elucidation of the mechanisms of early self-assembly and thermal decomposition of the archetypal MOF ZIF-8[48] (ZIF stands for Zeolitic Imidazolate Framework), that features many applications[49] and whose growth has been well studied from an experimental point of view.[23,26,28,29] ZIF-8 is formed by $Zn^{2+}$ cations tetrahedrally coordinated to 2-methylimidazolate ($MIm^-$) ligands forming a sodalite topology. Many synthesis routes are reported in the scientific literature for this material, both as a powder or as a membrane/film.[49,50] In what follows, we will concentrate on solvent-based syntheses. ZIF-8 has been synthesized in water,[51] methanol (MeOH),[52] Dimethylformamide (DMF),[48] and Dimethyl sulfoxide (DMSO),[53] among other solvents. Changing the solvent has been shown to modify the size and morphology of the obtained crystals,[53] which may influence the properties of the resulting material.[54] Besides the solvent, the metal to ligand ratio can also affect the result of the synthesis.[51] This plethora of synthesis works illustrates the complexity of ZIF-8's crystal growth.

In this contribution, we make a leap forward in the understanding of the self-assembly of ZIF-8 and its solvothermal decomposition. We shed light into molecular-level mechanistic details such as: (i) the nature and lifetimes of the different chemical species that can be found in solution at the early stages of self assembly, (ii) the role of the ligands as templating agents for the formation of sodalite cages, (iii) the formation of an intermediate highly connected amorphous aggregate and (iv) the height of the free energy barriers associated to the ZIF-8 crystal decomposition. Our results provide support to some mechanistic hypothesis that had been previously proposed to interpret experimental results,[55,56] and give further insight for the first time.

This paper is structured as follows. In Sec. II, we present the force field, its validation and the collective variables used in the MD simulations, along with a detailed discussion on why we chose them. Sec. III is devoted to the analysis and rationalization of the simulations results. Our main conclusions are summarized in Sec. IV.

## II. METHODOLOGY

### A. Simulation Details

The Large–scale Atomic/Molecular Massively Parallel Simulator (LAMMPS)[57] code was used to perform all the simulations. Energy optimizations were calculated using a self–consistent cycle at each step of the minimization, where the cell volume and shape were anisotropically relaxed using the Polak and Ribière[58] version of the conjugate gradient algorithm followed by an optimization of the atomic positions using the damped dynamics method described by Bitzek et al.[59] MD simulations were run in the NPT ensemble using the Nosé-Hoover thermostat with a timestep of $dt = 0.5$ fs. Thermostat and barostat relaxation times were set to $10dt$ and $100dt$, respectively. The Ewald summation method was used to calculate the electrostatic contribution to the energy.

The initial coordinates of the reactive species were generated as follows: a) NVT Monte Carlo (MC) simulations were performed using the RASPA code,[60] for those systems starting from a completely uncoordinated phase (that is the solvated metal cation and ligands), and b) the experimental crystallographic coordinates were subsequently geometry optimized via the LAMMPS code for systems starting from a crystalline phase. In both cases the solvent coordinates were generated via a subsequent NVT MC simulation. For the simulations starting from the crystalline phase, the dummy atoms were added to the structure by means of a in-house program written in FORTRAN. The initial atomic positions for the ZIF-8 crystal were taken from Park et al.[48] and a $2 \times 2 \times 2$ supercell was built from them. The LAMMPS inputs were automatically generated from the coordinates obtained from the NVT MC simulation by another in-house pro-gram written in FORTRAN. Both have been uploaded to a publicly available online repository: https://github.com/salrodgom/ZIFWTMetaDnbZIFFF. The WT-MetaD simulations were performed with the multiple walker enhanced sampling technique,[61] using 6 walkers that were run with 32 cores each.

Table I: Simulation experiments that were carried out in this work.

| Initial condition | Cell Size | Type | $T$ / K | Solvent | $Zn^{2+}$ & solvent numbers |
|---|---|---|---|---|---|
| ZIF-8 crystal | $2 \times 2 \times 2$ | Unbiased MD | 300, 450, 600, 800, and 1000 | MeOH | 96 and 400 |
| ZIF-8-like cluster | $\sim 68$ Å | Unbiased MD | 298 | EPS | 442 and 6864 |
| Solvated $Zn^{2+}$ and $MIm^-$ | $\sim 33$ Å | WT-MetaD | 300, 370, 450 | MeOH | 12 and 512 |
|  | $\sim 67$ Å |  |  |  | 96 and 2240 |
|  | $\sim 67$ Å |  |  |  | 96 and 4096 |
|  | $\sim 100$ Å |  |  |  | 324 and 13824 |
| ZIF-8 crystal | $2 \times 2 \times 2$ | WT-MetaD | 300, 450 | MeOH | 96 and 400 |
|  |  |  |  |  | 96 and 300 |
|  |  |  |  |  | 96 and 447 |

For the WT-MetaD simulations performed to study the polymerization of $Zn^{2+}$ and $MIm^-$ moieties, we tested systems of different sizes, namely 12, 96, and 324 $Zn^{2+}$ cations. In all simulations, the charge was neutralized with $MIm^-$ anions in a stoichiometric proportion. In these cases, we have kept the explicit solvent (MeOH) molecules/$Zn^{2+}$ ratio constant and equal to $128/3$. In addition, for the system with 96 $Zn^+$ cations, we have also studied a system with $\sim 55\%$ of this ratio. For the system starting from the ZIF-8 we loaded the structure with MeOH by performing $\mu$VT MC simulations at 1 bar and 298 K. Moreover, we have studied two lower loadings at standard conditions ($\sim 90\%$ and $\sim 66\%$).

Tab. I lists the simulation experiments that were conducted. Each WT-MetaD simulation took about 200 thousands CPU hours, and the overall computational cost of the whole project was of about 3 million CPU hours.

### B. nb-ZIF-FF force field development

To the best of our knowledge, Hu, Zhang, and Jiang [62] developed the first ZIF-8 force field, which was later updated.[63] Since then, several groups have developed new models to reproduce ZIF-8's experimentally observed properties, such as its mechanoelastic behavior, its infrared spectrum and the *flapping* of ligands in the pore windows, among others.[64–67] All these force fields treat the interaction between the coordination centers, namely the $Zn^{2+}$ cations, and the nitrogen N atom of the ligand with a bonded potential, which precludes modeling bond breaking and formation. Since taking into account the bond formation and breaking equilibrium is essential for studying self-assembly and decomposition processes, we have developed a new force field: nb-ZIF-FF (nb stands for non-bonded). The intramolecular, electrostatic and van der Waals interactions of our force field are based on the force field by Weng and Schmidt,[66] called ZIF-FF. We discarded all the bonds, angles and torsions that imply Zn–N contacts and replaced them by a Morse functional form, which is known to capture the coupled motion of coordination bonds.

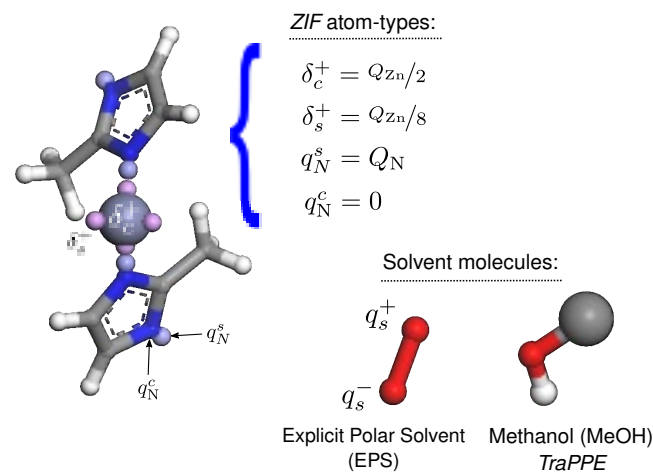

Figure 1: Scheme of the cationic and anionic dummy-atom models for $Zn^{2+}$ ($\delta_s^+$ and $\delta_c^+$ shell and core particles, respectively), $MIm^-$ ($N^s$ and $N^c$, shell and core particles for N atoms) as well as the two explicit solvent models that were considered in nb-ZIF-FF. N, C, H, Zn, O and the dummy particles are shown in blue, grey, white, light purple, red and pink/cyan, respectively. The Methyl group in MeOH is represented by a large gray sphere.

In addition, the spatial orientation of the $Zn^{2+}$—$N^-$ dipoles needs to be considered to achieve the correct ZIF-8 topology. For this, we relied on CDA models for the metallic cations $Zn^{2+}$, as well as for the $MIm^-$ ligand moiety (see Fig. 1). The CDA model,[34] as well as the electrostatic interactions between the different dummy atoms, will play a central role in modeling the directionality of the organic ligand assembly around the metal cation in the absence of angle and torsion interactions, as shown by Jawahery *et al.* [35] Within our CDA model, a central particle with charge $q_{Zn}^c$ is surrounded by four other charged particles, the shell dummies, with charges $q_{Zn}^s$, each of them placed in one of the four vertexes of a tetrahedron centered in the central particle. The distance between shell and central particles is 0.9 Å. The total sum of the charges is equal to the charge assigned to the $Zn^{2+}$ cation in the model by Weng and Schmidt:[66] $Q_{Zn} = q_{Zn}^c + 4q_{Zn}^s = 0.7072\ e$. While the central particle interacts *via* a sum of the Morse-based dispersion and Coulombic potentials, the shell atoms are only involved in electrostatic interactions. In addition, we have modeled the N atoms by a two sites anionic dummy atom model: a dummy particle containing all the N charge $Q_N = q_N^s$, was introduced at a distance of 0.5 Å from the original particle modeling N, which is located at the atom's position within the MIm$^-$ cycle and only interacts through dispersion interactions modeled *via* the Morse potential. The charge distribution was chosen so as to reproduce the radial distribution functions and other structural properties in crystalline ZIF materials. The original harmonic potential considered by Weng and Schmidt[66] for modeling bonded contributions has the electrostatic interaction embedded within, we took this into account when parameterizing our Morse potential. A detailed explanation of the potential fitting process, as well as some additional details related to the force field are given in the Subsec. II C and in the Supporting Information (SI), Sec. SII.

Even though nb-ZIF-FF was exclusively applied to model ZIF-8 in the context of this work, we note that it can also be applied to model other ZIFs, as discussed in the Subsec. II C. Indeed, we have calculated different structural and energetic properties of many ZIFs using nb-ZIF-FF and obtained excellent results compared to experimental and first principles calculations data as shown in Tab. II.

The solvent was treated *via* three different force field models: an implicit and two explicit ones. Implicit solvent models allow to sample much longer MD simulations times,[68] while an explicit solvent is necessary to study the collaborative mechanisms between solvent and reactants during the nucleation process.[32,33] For the implicit solvent, we have employed a corrected Coulomb interaction in which the dielectric permittivity, $\varepsilon$, is distance dependent. At short distances the relative permittivity tends to a dielectric saturation value $\varepsilon_c$ —close to vacuum—, while at long distances the relative permittivity reaches its bulk limit: $\varepsilon_r \to 78.2$, which corresponds to the value for water. To model the solvation shell around the inorganic and organic moieties, a small potential barrier has been added to frustrate a certain number of attempts of metal–ligand approaches, thus mimicking the role of the solvent in the early stages of nucleation, as in the work by Nguyen and Grünwald.[68] We have used the implicit solvent model in preliminary work to set up the input files to study the WT-MetaD simulations and to explore the behavior of different collective variables at a reduced computational cost. All simulation results that we report come from the explicit solvent simulations. Most of these have been carried out by considering the united atoms TraPPE[69] force field model for Methanol (MeOH). The interatomic cross interactions between the solvent and the nucleating species were modeled by LJ potentials whose parameters were calculated by applying the Lorentz-Berthelot mixing rules. The third and final solvent model considered only for force field validation calculations was a simple dipole proposed by Biswal and Kusalik,[33] labeled EPS (from Explicit Polar Solvent) in Tab. I (see Fig. 1).

### C. nb-ZIF-FF Validation

nb-ZIF-FF has been carefully validated by testing it against a set of relevant structural and energetic properties compared with those obtained from experimental and first principles calculations data, namely: *a*) the structural stability of the crystalline geometries of a series of ZIFs as well as of *b*) a solvated nanocluster of ZIF-8 at 298 K and *c*) the structural and geometric parameters as well as *d*) the elastic constants for a series of ZIFs.

Dürholt *et al.*[67] and Lewis *et al.*[70] have computed the energies of a series of highly stable ZIFs *via ab initio* calculations. They have predicted that the energies follow the trend: $E_{zni} < E_{cag} < E_{BCT} < E_{MER} < E_{SOD} < E_{DFT} < E_{GIS}$. We computed the energy relative to $E_{zni}$ for the same ZIFs subset using nb-ZIF-FF, and found that the trend is quite well reproduced: $E_{zni} < E_{cag} < E_{BCT} < E_{DFT} < E_{SOD} < E_{MER} < E_{GIS}$. A table with the energies can be found in the SI (Tab. S2). The most significant difference is in the relative energy of the DFT structure. We believe that this is due to the fact that DFT presents two types of Zn-Ligand-Zn angles that nb-ZIF-FF is not able to resolve, since it has no bending terms to account for them. This is not a problem for our current work, because we are using this force field to model the self-assembly of ZIFs and not the crystalline materials themselves. Because most of these structures are composed by Imidazolate ligands (instead of MIm$^-$), we added some new atom types to the force field to adapt it. These are minor changes, that do not modify the force field in its essence, more information can be found in the SI (Sec. SII).

To further probe the nb-ZIF-FF force field, we determined the cell parameters for the series of ZIFs mentioned above as well as for ZIF-8 and compared them to experimental, *ab initio*- and force field-based calculations data. The results are presented in the SI, Tab. S1. We observe a good agreement between the nb-ZIF-FF calculated and experimental cell parameters. Our calculations are also consistent with both previously published force field- and electronic structure-based calculations.[67,70]

We further put our force field to test by assessing the structural stability of a ZIF-8 nanocluster (see Fig. 2a), for which the surface exposed to the solvent represents roughly 25% of the material. The cluster was built by cropping a periodic $3 \times 3 \times 3$ ZIF-8 structure and it was inserted into a $a = 100$ Å cubic simulation box. Because the clean cuts of periodic systems lead to unrealistic surfaces, some extra surface elements (MIm$^-$ and $Zn^{2+}$) were added to minimize the surface electric charges. The cluster was immersed in the EPS solvent proposed by Biswal and Kusalik,[33] such that the system contained 442 $Zn^{2+}$, 884 MIm$^-$ and 6864 EPS molecules. The snapshot shown in Fig. 2a was collected after 5 ns of MD simulation at 298 K and it shows that the sodalite cages of the ZIF-8 nanocluster maintain their shape. Interestingly, we observed that some MIm$^-$ ligands at the cluster surface are dissociated and adsorbed and some $Zn^{2+}$ cations are almost completely surrounded by solvent molecules. The same behavior was observed during the self-assembly simulations that will be detailed in Sec. III.

Table II: Cell parameters of a series of highly stable ZIF structures determined through geometry optimizations using nb-ZIF-FF (upright bold) and reference data from experimental (italic) and theoretical works.

|  |  | $a$ / Å | $b$ / Å | $c$ / Å | $\alpha$ / ° | $\beta$ / ° | $\gamma$ / ° |
|---|---|---|---|---|---|---|---|
| *zni* | **nb-ZIF-FF** | **23.28** | **23.28** | **12.64** | **90.01** | **90.00** | **90.00** |
|  | MOF-FF[67] | 23.23 | 23.23 | 12.79 | 90.00 | 90.00 | 90.00 |
|  | DFT[67] | 23.35 | 23.35 | 12.56 | 90.00 | 90.00 | 90.00 |
|  | *Exp.*[71] | *23.50* | *23.50* | *12.46* | *90.00* | *90.00* | *90.00* |
| BCT ZIF-1 | **nb-ZIF-FF** | **9.96** | **15.30** | **15.13** | **90.00** | **97.25** | **90.00** |
|  | MOF-FF | 10.09 | 14.55 | 15.91 | 90.00 | 117.00 | 90.00 |
|  | DFT[70] | 9.74 | 15.39 | 15.18 | 90.00 | 98.55 | 90.00 |
|  | *Exp.*[48] | *9.74* | *15.26* | *14.93* | *90.00* | *98.62* | *90.00* |
| DFT ZIF-3 | **nb-ZIF-FF** | **18.03** | **18.54** | **13.47** | **102.37** | **89.93** | **89.95** |
|  | DFT[70] | 18.96 | 18.96 | 16.77 | 90.00 | 90.00 | 90.00 |
|  | *Exp.*[48] | *18.97* | *18.97* | *16.74* | *90.00* | *90.00* | *90.00* |
| MER ZIF-10 | **nb-ZIF-FF** | **26.92** | **26.92** | **20.16** | **90.00** | **90.02** | **89.78** |
|  | DFT[70] | 25.96 | 25.96 | 19.35 | 90.00 | 90.00 | 90.00 |
|  | *Exp.*[72] | *27.37* | *27.37* | *18.58* | *90.00* | *90.00* | *90.00* |
| GIS ZIF-6 | **nb-ZIF-FF** | **19.34** | **19.34** | **20.14** | **90.00** | **90.00** | **90.00** |
|  | MOF-FF | 19.23 | 19.23 | 19.80 | 90.00 | 90.00 | 90.00 |
|  | DFT[70] | 18.57 | 18.57 | 21.00 | 90.00 | 90.00 | 90.00 |
|  | *Exp.*[48] | *18.52* | *18.52* | *20.25* | *90.00* | *90.00* | *90.00* |

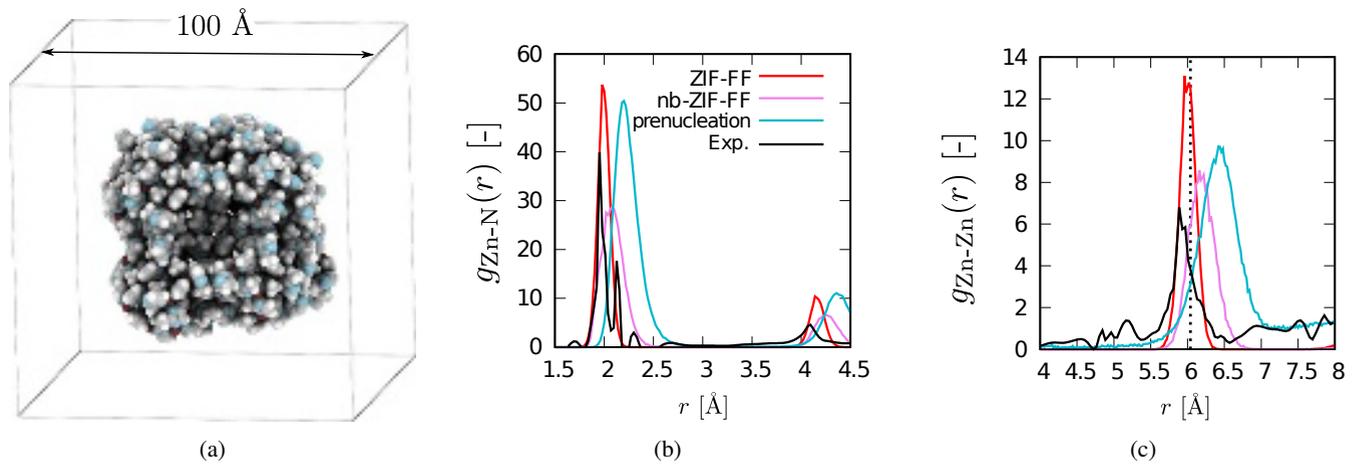

Figure 2: (a) Snapshot of a ZIF-8 cluster. The core of the cluster remains stable but the connectivity of MIm$^-$ and Zn$^{2+}$ moieties at the surface changes over time. (b) Zn-N and (c) Zn-Zn radial distribution functions $g(r)$. ZIF-FF simulated, nb-ZIF-FF simulated and experimental results (Refs. 73,74) are shown in red, pink and black lines respectively. The cyan solid lines show the corresponding $g(r)$ of an under-coordinated cluster formed during the polymerization simulations.

We also computed the pair–correlation functions $g(r)$ between Zn–N and Zn–Zn atoms. Fig. 2b and 2c show that there is a good agreement between nb-ZIF-FF predictions, those by ZIF-FF and the experimental values measured by Cao *et al.*[73] and Bennett *et al.*[74] The distances of the first maximum and minimum for Zn-N are 2.0 Å and 3.0 Å, respectively, and for Zn-Zn pairs they are 6.0 Å and 7.5 Å respectively. Pair correlation functions during a pre-nucleation stage from a nb-ZIF-FF unbiased MD simulation in the canonical ensemble at 298 K are also shown (cyan line in Fig. 2b and 2c). The first peak is displaced to larger distances for low degrees of polymerization (which can be correlated to less Zn–N bonds and more MeOH–Zn and MeOH–N interactions). This highlights the central role of the solvent in the self-assembly process. In addition, the Zn–N $g(r)$ corresponding to several other ZIFs crystal structures are reported in the SI (see Fig. S5). All of them show values that are in good agreement with the experimentally-measured ones.

As a final stringent validation criterion, we computed the elastic constants for our selected series of ZIFs. SOD ZIFs results are presented in Tab. III while those for the rest of the ZIFs are shown in the SI. The agreement between our elastic constant calculations and the experimental/calculated values is excellent.

Table III: Elastic constants of ZIF-8 and SALEM-2 computed with nb-ZIF-FF (upright bold), other force field-based or first principles calculations and experimental measurements (italic).

|  |  | $C_{11}$ / GPa | $C_{12}$ / GPa | $C_{44}$ / GPa |
|---|---|---|---|---|
| ZIF-8 | nb-ZIF-FF (this work) | **12.67** | **4.00** | **2.915** |
|  | ZIF-FF (this work) | 9.33 | 6.45 | 1.37 |
|  | MOF-FF (by Dürholt et al.[67]) | 8.54 | 6.55 | 0.62 |
|  | DFT (by Tan et al.[75]) | 11.038 | 8.325 | 0.943 |
|  | Exp. (by Tan et al.[75]) | *9.523* | *6.865* | *0.967* |
| SALEM-2 | nb-ZIF-FF (this work) | **6.333** | **3.775** | **2.6204** |
|  | DFT (by Zheng et al.[76]) | 8.949 | 7.59 | 2.36 |
|  | MOF-FF (by Dürholt et al.[67]) | 6.65 | 4.95 | 1.12 |

All of the above validation tests give us confidence to pursue the study of the self-assembly of ZIF-8 considering nb-ZIF-FF as a reliable force-field.

### D. Sampling of nucleation and decomposition events

Since MOF nucleation and thermal decomposition are activated processes, the events that lead to them are rarely sampled in a molecular dynamics simulation and the ergodic hypothesis is not fulfilled. In this work, we tackle this problem by relying on the multiple-walkers Well-Tempered Metadynamics[43,61] enhanced sampling method. To briefly illustrate how WT-MetaD works, let us consider a couple of functions of the atomic coordinates that adequately describe the nucleation process in a low-dimensional space. These functions are usually called collective variables (CVs), $s(q)$, and they allow to distinguish between the different states associated to the local free energy minima that characterize the physically relevant portion of the free energy surface. In WT-MetaD,[43] a history-dependent bias potential that drives the system evolution along the CVs is added to the Hamiltonian. This potential is defined such that energy penalties in the form of Gaussians are periodically deposited over the free energy surface in the CVs space, thus reducing the probability of revisiting configurations that have already been sampled. The rate of deposition of this bias potential decreases over the simulation time. The probability distribution sampled by the biased system is given by:

$$P_{\text{WT-MetaD}}(s(q)) \propto P(s(q))^{\frac{1}{\gamma}} \quad (1)$$

where $P(s(q))$ is the probability distribution sampled in the isothermal–isobaric ensemble and $\gamma = T+\Delta T/T$. $\gamma$ is chosen as an estimate of the height of the free energy barrier between the relevant states.[44] Here, we estimated the height of the barrier based on the results of an *in situ* XRD study performed by Cravillon et al.[77] The authors estimated the nucleation rate constants in the 393 K to 413 K temperature range by fitting crystallization data with the Gualtieri equation for solvothermal synthesis carried out in MeOH. They subsequently estimated the activation energy of nucleation from an Arrhenius plot $E_a = 69$, then $\gamma(400K) = \frac{8300K}{400K} \sim 20$. We thus chose $\gamma = 20$ for our simulations at $T = 400$ K. We also explored other values, namely $\gamma = 3, 10, 40$. We considered a height of $2k_BT$ for the deposited Gaussians. The width was estimated for each CV by running unbiased MD simulations and computing the standard deviation of the CVs distribution.

CV selection can be very challenging as a poor choice of CVs can prevent the free energy calculation from converging.[44] For this reason, we decided to systematically explore different CVs, including the simulation box volume, the enthalpy, the connectivity and local- and long-range- order parameters. Two connectivity-related CVs were considered: $\kappa_1^c$ and $\kappa_2^c$ where $c$ stands for connectivity. The normalized coordination $\kappa_1^c$ is the number of Zn—N bonds relative to the total number of bonds, which is equal to the number of Zn atoms times 4 for perfect tetrahedral connectivity, $4N_{\text{Zn}}$, as would be the case in perfectly coordinated bulk ZIF-8 crystals:

$$\kappa_1^c = \frac{1}{4N_{\text{Zn}}} \sum_{\substack{i \in \text{Zn atoms} \\ j \in \text{N atoms}}} \sigma_{ij} \quad (2)$$

$\sigma_{ij} = \sigma(r_{ij})$ is a switching function dependent of $r_{ij}$ distance —*i.e.* equal to one when the corresponding Zn-N pair is connected and zero otherwise—. In order to make sure that the CV has continuous derivatives, we considered $\sigma_{ij} = \exp\left(-(r_{ij}-d_0)^2/2r_0^2\right)$ as implemented in PLUMED with $d_0 = 2.35$ and $r_0 = 0.2$ (see Code 1 in Sec. SI of the SI).[78,79] $\kappa_1^c$ is zero when none of the $Zn^{2+}$ cations and ligands are coordinated —*i.e.* in the initial stage of $Zn^{2+}$ and $MIm^-$ mixing— and it takes the value of 1 for a fully-coordinated ZIF-8 crystal.

The second connectivity variable, $\kappa_2^c$ is related to the clustering, it represents the ratio of $Zn^{2+}$ cations belonging to the largest connected cluster (see Code 2 in Sec. SI of the SI). A cluster is defined as a continuous network formed by interconnected $Zn^{2+}$ and $MIm^-$ ligands. As such, this CV takes a value of zero at the initial stage of $Zn^{2+}$ and $MIm^-$ mixing and one when all the $Zn^{2+}$ cations belong to the same cluster. Both $\kappa_1^c$ and $\kappa_2^c$ were successfully used as CVs in an article by Tribello et al.[80]

We have also considered a subset of relevant order parameters. We considered the environmental similarity CV used by Piaggi and Parrinello,[81] which is based on the definition of local density of neighbors $j$ within an environment around a central atom $i$ and it is a generalized version of that proposed by Bartók, Kondor, and Csányi.[82,83] This CV measures how an atomic environment $\chi$ overlaps with a reference environment $\chi_0$ with the following kernel function:

$$\kappa_{\chi_0}(\chi) = \int d\mathbf{r}\rho_\chi(\mathbf{r})\rho_{\chi_0}(\mathbf{r}) = \langle \rho_\chi | \rho_{\chi_0} \rangle \propto$$
$$\propto \sum_{\substack{i \in \chi \\ j \in \chi_0}} \exp\left(-\frac{|\mathbf{r}_i - \mathbf{r}_j^0|^2}{2\sigma^2}\right) \quad (3)$$

where $\rho_\chi$ is the density around a central $i$–atom and considering only the neighbors in a predefined environment $\chi$. This quantity is measured for all $Zn^{2+}$ cations, for $\chi$ and $\chi_0$, instantaneous and reference environments, respectively. For the ZIF-8 crystal thermal decomposition simulations, we have chosen the optimized positions of the Zn atoms at the ZIF-8 structure (*sod* topology) as the reference environment $\chi_0$. The *global* parameter, $\omega_{\text{sod}}^0$, for the whole system is constructed by calculating the ratio of Zn atoms with $\kappa_{\text{sod}}(\chi) > \kappa^0$ (see Code 3 in Sec. SI of the SI and Fig. S2). For the simulations starting from the solvated $Zn^{2+}$ and $MIm^-$ moieties, the reference environment $\chi_0$ was set as the first Zn-Zn tetrahedral environment (see Code 4 in Sec. SI of the SI). In the same way than $\omega_{\text{sod}}^0$, $\omega_{\text{tet}}^0$ is constructed as the ratio of Zn atoms than fulfill the reference tetrahedral environment (highlighted in Fig. S2).

Finally, we tested the order parameters that were introduced by Steinhardt, Nelson, and Ronchetti.[84] In essence, the orientational Steinhardt parameters, $\{q_{lm}(\mathbf{r}_i)\}$, can be understood as the projection of the "near neighbours" $\sigma_\chi$ on the spherical harmonics around ($r < r_c$) the central $i$-atom, for a generic $l$ degree. This magnitude is an orientational order parameter and it is not invariant under rotation, so it cannot be used as such. To deal with this issue, ten Wolde, Ruiz-Montero, and Frenkel[85] introduced the local-averaged Steinhardt parameters,

$$\overline{Q_l}(i) = \frac{1}{N_\chi(i)} \sum_{j \in \chi} \sigma(\mathbf{r}_{ij}) \sum_{m=-l}^{+l} q_{lm}^*(i) q_{lm}(j) \quad (4)$$

which measure how many of the atoms around the $i$-atom have a similar pattern, $q_{lm}$ in their coordination spheres. The local-averaged Steinhardt parameters are invariant under rotations and are recommended for distinguishing between liquid and solid phases.[85,86] Their value is large for atoms immersed in a crystalline phase and small for those belonging to the solvated ions phase. We have calculated the average of these parameters for all Zn atoms for $l = 3, 4,$ and $6$ as implemented in the PLUMED code (v2.7).[78] We have finally selected $\omega_{Q6}^o$, the ratio of Zn atoms with $\langle \overline{Q_l} \rangle > \langle \overline{Q_l}(\text{sod}) \rangle$ as a biasing CV (see Code 5 in Sec. SI of the SI).

## III. RESULTS AND DISCUSSION

### A. Selection of Collective Variables

We seek to study two transformations, one between the solvated $Zn^{2+}$ and $MIm^-$ moieties and the aggregate they form after polymerization (early stages of ZIF-8 nucleation) and the other between the ZIF-8 crystal (Fig. 3a right) and the non-crystalline phase that is obtained as a result of thermal decomposition at $T > 600$ K (Fig. 3a, left). In order to circumvent the energetic barriers that characterize these transformations we selected three CVs for our WT-MetaD simulations for each system. Two of them were used for both transformations: $\kappa_1^c$, the Zn-N connectivity ratio and $\omega_{Q6}^o$, the ratio of Zn atoms with $\langle \overline{Q_6} \rangle > 0.8$ (see Fig. S3). The third CV was $\omega_{\text{sod}}^o$, the ratio of Zn atoms with $\chi_{\text{sod}} > 0.7$ for the ZIF-8 crystal decomposition simulations and $\omega_{\text{tet}}^o$, the ratio of Zn atoms with $\chi_{\text{tet}} > 8$ for the simulations of the early polymerization of $Zn^{2+}$ and $MIm^-$ moieties. These CVs were selected so that one of them is strongly correlated with enthalpy ($\kappa_1^c$) and the other two are related to order.

The time evolution of the two connectivity related CVs ($\kappa_1^c$ and $\kappa_2^c$) during the first nanosecond of a WT-MetaD simulation starting from the uncoordinated $Zn^{2+}$ and $MIm^-$ solvated moieties is shown in Fig. S1. We found that these two CVs are highly correlated among themselves as well as with the enthalpy and the simulation box volume, so we decided to keep only $\kappa_1^c$ as a representative for the connectivity/enthalpy group.

The order-related CVs that were selected for the early polymerization simulations describe different aspects of local ordering. On the one hand, $\omega_{Q6}^o$ describes the symmetry of the arrangement of the first neighbor shell and also from the closest neighboring atoms beyond the tetrahedron, it is based on spherical harmonics and not in reproducing a particular reference configuration. Among the three Steinhardt parameters–based CVs, we have selected $\omega_{Q6}^o$ instead of the $Q_3$- and $Q_4$–based CVs because it is the most commonly used in the literature to characterize solid/liquid states. On the other hand, $\omega_{\text{tet}}^o$ focuses on reproducing the $Zn(MIm)_4^{2-}$ tetrahedron configuration by construction. For the ZIF-8 crystal decomposition study we kept $\omega_{Q6}^o$ because of its generic character and its ability to describe local order, but the targeted-structure CV $\omega_{\text{tet}}^o$ was replaced by $\omega_{\text{sod}}^o$ because in this case we aim to reproduce the sodalite crystal topology as reference configuration, and not only the tetrahedron formed by the first neighbors. In this sense, $\omega_{\text{sod}}^o$ could be interpreted as a long-range order CV.

A basic pre-requisite for a CV to be appropriate for exploring the free energy surface that underlies a chemical transformation is that it needs to adopt different values for the states that are to be identified. To verify that this pre-requisite is fulfilled, we calculated the distribution of our chosen CVs for the different states described above. Fig. 3b shows these distributions for the three selected CVs used to study the ZIF-8 crystal decomposition process. At low temperatures, the system exhibits the ZIF-8 crystal phase, while at high temperatures ($T > 600$ K) it forms a non-crystalline phase. It is clear that the three CVs can differentiate the two states quite well. We established cutoff values for the CVs to define whether the system is crystalline or not in a binary fashion. These cutoffs are illustrated both in Fig. 3b and Fig. S3 for the selected CVs and for some of the remaining CVs respectively. We have also performed the same analysis for simulations starting from the uncoordinated metal ions and ligands phase. All distributions are centered on zero at the beginning of the simulations and $\kappa_1^c$ quickly rises as the nucleation process progresses.

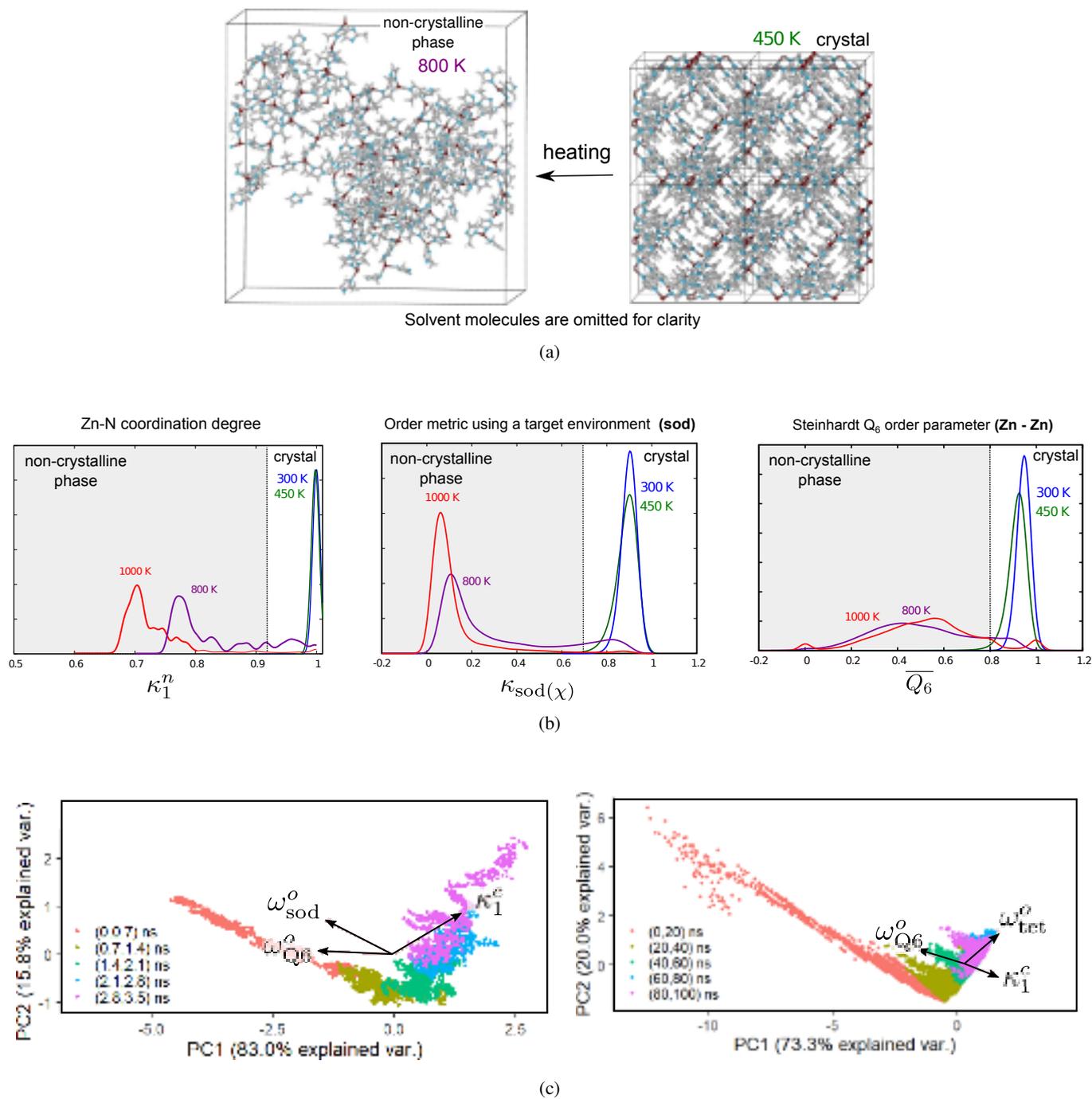

Figure 3: (a) Two snapshots taken from unbiased NPT MD simulations at (Left) $T = 800$ K, and (Right) $T = 450$ K. (b) Probability density functions for some CVs at 300–1000 K: (Left) Zn-N coordination degree, (Middle), environmental similarity $\kappa_{\text{sod}}(\chi)$ (critical value of $\kappa^0 = 0.7$), and (Right) Steinhardt $\langle \overline{Q_6} \rangle$ parameter (critical value of $\langle \overline{Q_6^0} \rangle > 0.8$). (c) First two components of the principal component analysis performed on the values of all the explored CVs harvested along an unbiased MD simulation at $T = 800$ K starting from the ZIF-8 crystal (left) and $T = 298$ K starting from the uncoordinated solvated $Zn^{2+}$ and $MIm^-$ moieties (right). Both figures incorporate the projection of the selected CVs into the PC1-PC2 space.

To make sure that the selected CVs are not redundant, we run unbiased MD simulations for $T = 300, 450, 500, 600, 800$, and $1000$ K lasting 3-4 ns in the NPT ensemble starting from a ZIF-8 crystal configuration filled with MeOH, and we calculated the values of the above mentioned CVs along the simulation trajectory, as well as of volume and enthalpy. Another MD simulation was carried out at 298 K starting from the uncoordinated MIm$^-$ ligands and Zn$^{2+}$ cations immersed in MeOH for 100 ns to characterize the initial stages of the self-assembly process.

We then built a dataset combining all the harvested CV values and performed Principal Component Analysis (PCA). PCA consists on finding a new basis of orthogonal vectors in which the data are represented. These vectors $\mathbf{v}_i$ are called the principal components, they are the eigenvectors of the covariance matrix and they are organized in such a way that the total variance of the data represented by vector $\mathbf{v}_n$ is larger than that covered by vector $\mathbf{v}_{n+1}$. It is possible to profit from this property to choose only the first few principal components to represent the data, since they will allow to account for the largest part of the variance that characterizes it. By performing PCA over the CVs values along the simulation, we can see how the system evolves in the PC space. Each point in the PC space represents the state of the system at a particular time. The euclidean distance between two points in the PC space can be used as a proxy for measuring how drastic has been the system's evolution between them. In addition, it is possible to measure the contribution of a particular CV to the evolution of the system in the PC space by computing the projection of the CV vector in the PC space.

The results of these analyses are shown in Fig. 3c and Fig. S6, where each point represents the position of the system in the PC1-PC2 space (the two principal components that best represent the dispersion of the data). At low temperature ($T = 300$ K), we found that the system does not explore the PC space: the points are uniformly distributed in a small region of the PC space, and their distribution does not change with simulation time (see Fig. S6). This observation is consistent with the direct inspection of the trajectory, which shows a stable ZIF-8 crystal all along the simulation. The degree of changes suffered by the system gradually increases with increasing temperature. The most drastic changes happen at $T > 600$ K. Indeed, this also correlates with the visual inspection of the trajectory that shows that at higher temperatures, more MIm$^-$ and Zn$^{2+}$ moieties become dissociated, and at the limit of $T > 600$ K we recover a non-crystalline phase, as shown in Fig. 3a left.

If we turn our attention to the evolution of the system in the PC space with time, we see that for the ZIF-8 crystal decomposition at high temperature (Fig. 3c left panel) there is an initial transient period where the system suffers drastic changes (moves considerably along the PC space), mostly driven by the change in the cell volume produced by thermal expansion. After this period, which is quite short ($\sim 500$ ps), the system evolution is drastically slowed down. This latter stage corresponds to the formation of a non-crystalline phase. We note that temperature alone, with the consequent changes in the volume, is enough to drive the decomposition of the ZIF-8 crystal. The projection of our chosen CVs on the PC1-PC2 space are represented by vectors that point to different directions in the PC space, thus proving that the selected CVs are not redundant as they allow to push the system into different realms in the PC space. The same behavior is found for early polymerization simulations starting from the solvated MIm$^-$ and Zn$^{2+}$ moieties (see Fig. 3c right panel). Here, the selected CVs are also non redundant and we can also observe a fast initial period when the system undergoes important changes, associated with the polymerization process and the formation of a highly-connected amorphous phase. In what follows, we will call the two non-crystalline phases that we obtain from the self-assembly simulations and the thermal decomposition simulations amorphous and vitreous phase, respectively.

### B. Free energy surface calculations

#### 1. Early polymerization of MIm$^-$ and Zn$^{2+}$ moieties

We further implemented our selected CVs into WT-MetaD simulations on the NPT ensemble starting from the solvated Zn$^{2+}$ and MIm$^-$ moieties to study the free energy surface and mechanistic details of the initial polymerization stages of the ZIF-8 self-assembly.

We have performed a very short equilibration (less than 1 ps) because Zn-N bonds start forming early in the simulation as well as the first polymerization events, such as the formation of Zn(MIm)$_2$ and Zn(MIm)$_3^-$ clusters. This is consistent with the observations by Filez *et al.*[30] concerning the early stages of formation of ZIF-67.

In the first 20 ps $\sim 50$% of all possible Zn-N bonds have already been formed, and 85% of them within the first 150 ps. For $t > 150$ ps, $\kappa_1^c$ slowly increases until it reaches a plateau for a value that depends on the size of the system (see Fig. S1). This behavior can be ascribed to the fact that the size of the system is key to achieve the formation of a 3D highly-connected amorphous aggregate. For the same composition (in terms of Zn atoms), if the total amount of solvent is less than a critical amount, percolated 1D, 2D or 3D-linked clusters are formed. Fig. 4 illustrates the clusters that are formed in our WT-MetaD simulations. The low dimensionality clusters (rods, chains or planes) persist over time along all three simulation setups shown in Fig. 4 (12 Zn(MIm)$_2$ 512 MeOH, 96 Zn(MIm)$_2$ 4096 MeOH and 324 Zn(MIm)$_2$ 13824 MeOH). The formation of low-dimensionality clusters has also been observed in previous simulation works[31,32] and could be in line with the idea of *aufbau* self-assembly mechanisms,[88] which postulate that 1D chains are first formed and associated to yield 2D planes that finally aggregate into 3D networks. Some of the clusters in Fig. 4 percolate through the periodic boundary conditions, which is not desirable at this stage of the study as it implies an artificial stabilization of the system. Indeed, such percolation phenomena may be favored only because they lead to a huge drop in potential energy. To avoid nonphysical effects, we work with relatively high $^{MeOH}/_{Zn(MIm)_2}$ ratios. However, a realistic study of the noncrystalline-to-crystalline transition must consider percolation through realistic periodic boundary conditions, that cannot be achieved via an atomistic model such as the one we employ here due to restrictions in the cell size to perform the calculation within a reasonable computing time. It is thus necessary to work with larger systems, for example using coarse grained models and within a simulation ensemble that allows the number of solvent molecules to change, such as the constant chemical potential ensemble used by Karmakar et al.[47] Indeed, solvent-reactants interactions are crucial during the different stages of self-assembly, and large variations of local solvent densities at the different stages of the self-assembly process are to be expected. The dynamics and long-range connectivity of these low-dimensionality clusters will be further explored in future work combining in situ spectroscopy experiments with multiscale modeling techniques.

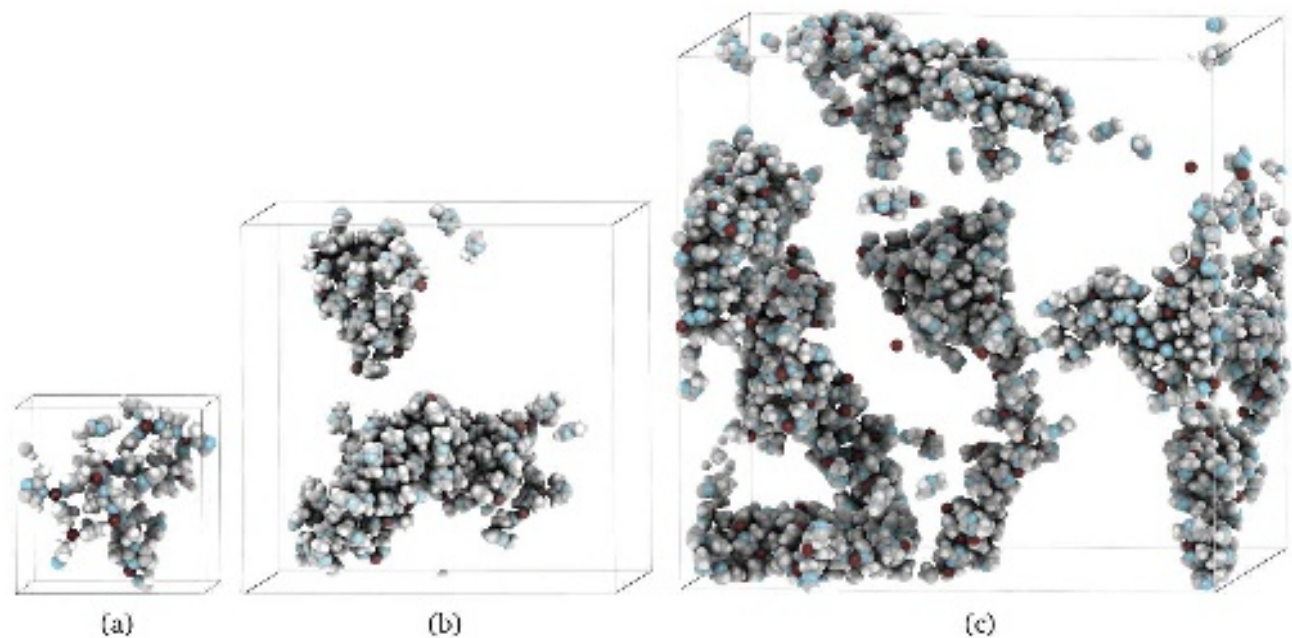

Figure 4: Snapshots depicting the amorphous aggregate phases formed in systems with compositions: (a) 12 Zn(MIm)$_2$ 512 MeOH, (b) 96 Zn(MIm)$_2$ 4096 MeOH, and (c) 324 Zn(MIm)$_2$ 13824 MeOH. The $^{MeOH}/_{ZnMIm_2} = {}^{128}/_3$ ratio is kept constant. Clusters were generated in WT-MetaD simulations at $T = 400$ K using $\kappa_1^c$, $\omega_{sod}^o$, and $\omega_{Q6}^o$ as CVs and the snapshots were taken at $t = 20$ ns using the iRASPA software.[87] Simulation boxes are in the same scale. Dummy atoms and solvent molecules are omitted for clarity.

We have further studied the evolution of the clusters that are formed during these first polymerization stages of the self-assembly process. In the first 50 - 200 ps there is a competi-tion between different clusters to be the largest. The number of competing clusters depends on the size of the system. Clusters are created/destroyed by either one of two mechanisms: (i) direct aggregation/breakage of clusters or (ii) evaporation/condensation of moieties into/from other clus-ters. For low $T$ systems ($T = 298; 370$ K with composition 96 Zn(MIm)$_2$ 4096 MeOH the smaller clusters evaporate after 200 ps and the remaining free Zn$^{2+}$ and MIm$^-$ moieties merge with the largest cluster (see Fig. S18). However, the main growth mechanism is the direct aggregation of clusters. At $T = 450$ K and 96 Zn(MIm)$_2$ 2240 MeOH composition, the smaller clusters do not evaporate quickly, and some small clusters coexist alongside the larger cluster —i.e. $\kappa_2^c < 1$ (see Fig. S16 and Fig. S19(g-l), for free energy surfaces -FES- and representative snapshots). The dependence of $\kappa_2^c$ on $T$ and composition is not clear, elucidating it will be the object of future research. The dominant clusters that are formed exhibit an amorphous aggregate topology, that could be assimilated to the amorphous highly-connected phase previously postulated as an intermediate species in the ZIF-8 self-assembly mechanism,[55,56] and that was also observed in the joint experimental/modeling work by Filez et al. for ZIF-67. This type of network is very resilient, its general appearance remains unchanged along the WT-MetaD simulations. Indeed, after 10-30 ns the local order-related CVs do fluctuate, but their mean values remain constant, and the nucleation process is stuck. This result may be an artifact of the simulation conditions, as the crystallization process depends on the local concentration of reactants and solvent molecules near the nuclei interfaces. Since our simulations are run in the NPT ensemble (constant number of molecules), the excess of surface solvent may hinder the system's ability to crystallize further.

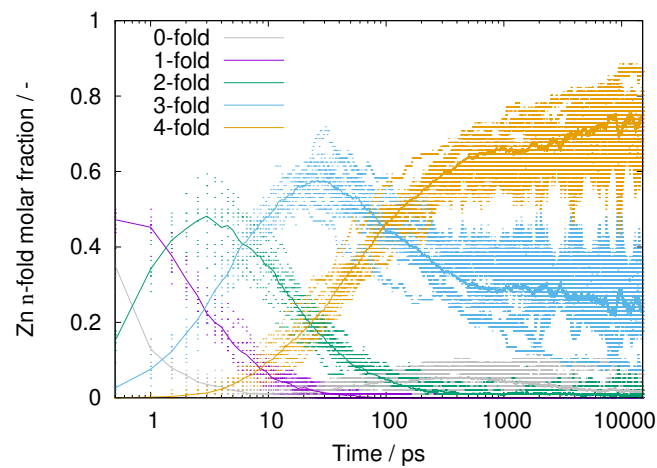

Figure 5: Population of $n$-fold coordinated Zn$^{2+}$ cations in the simulation box over time for the 96 Zn(MIm)$_2$ 4096 MeOH system from WT-MetaD simulations over 12 walkers (same system as in Fig. 4b).

As a continuation of our study, we analyzed the different species present in solution that drive the early stages of the nucleation process, the so-called pre-nucleation building units (PNBUs). Note that in our simulations Zn$^{2+}$ cations tend to be tetracoordinated; which is not surprising since the force field is optimized for that purpose by using the CDA model, although the presence of tricoordinated cations is not negligible (see Fig. 5). This is partly explained by the existence of surfaces that necessarily lowers the coordination of all species that constitute them, but also by the fact that the simulation is biased to explore different values of the Zn-N connectivity in the WT-MetaD (via the $\kappa_1^c$ CV) and the cluster surface/volume ratio. In Fig. S17 we show the 2D-FES using as CVs the mean connectivities Zn-N (nitrogen atom of MIm$^-$) and Zn-O (the oxygen atom of methanol). A clear trend is observed as methanol is displaced by the MIm$^-$. We also observe the presence of free 4-fold connected Zn(MIm)$_4^{2-}$ and 3-fold connected Zn(MIm)$_3^-$ with half-lives of $\sim 10$ ps, and free MIm$^-$ anions with a half-life of $\sim 1$ ps. Zn(MIm)$_x$ with $x < 3$ are very rare for simulation times greater than 100 ps. Larger PNBUs corresponding to rings are also formed. We observed the formation of 4-membered rings comprising both tetracoordinated and tri-coordinated metal cations. These rings are quite stable with half-lives of several nanoseconds ($\sim 10$ ns). 5- and 6-membered rings are also present, both having similar stability ($\sim 1$ ns). Direct inspection of the trajectory indicates that free MIm$^-$ moieties act as "templating agents" in the formation of cages from the rings.

Finally, we computed the free energy surface in the CVs space for the early polymerization of Zn$^{2+}$ and MIm$^-$ moieties, results are discussed in the next section.

#### 2. ZIF-8 crystal thermal decomposition

As a continuation of our work, we probed the influence of the amount of adsorbed solvent in the ZIF-8 crystal decomposition process via WT-MetaD simulations using $\kappa_1^c$, $\omega_{\text{sod}}^o$, and $\omega_{Q_6}^o$ as CVs.

While the exploration of the two order-related CVs is efficient, with frequent barrier crossings observed during the whole WT-MetaD simulation, test simulations with 100% of loading at standard conditions show that once small enough connectivity values are reached ($\kappa_1^c \sim 0.7$), the system does not come back to its initial non-connected state during 500 ns of simulation. By performing a restrained WT-MetaD (with $\kappa_1^c > 0.7$) at 90% methanol loading at standard conditions, we observe, however, frequent recrossing of the three biased CVs (see Fig. S7 and Fig. S8). Indeed, as reported in Fig. S8 the system can visit microstates with low CV values for $\omega_{\text{sod}}^o$ and $\omega_{Q_6}^o$, and it can recover its initial state further along the simulation. The connectivity CV, $\kappa_1^c$, is well explored in the restricted [0.7,1] interval. Indeed, we believe that the decomposition could not proceed up to the uncoordinated Zn$^{2+}$ and MIm$^-$ solvated moieties because of the limitations imposed by the fixed amount of solvent molecules during the simulation, as previously discussed. However, we observed that while the unbiased $\omega_{\text{tet}}^o$ CV is almost always within the range $0.5 < \omega_{\text{tet}}^o < 1$ along the trajectory, it occasionally explores $\omega_{\text{tet}}^o < 0.5$ regions (see Fig. S8), which corresponds to an increase in the volume $V$ (see Fig. S12) and $\kappa_2^c < 1$, which means that the framework is not fully-connected (see Fig. S11). This leads to highly-disordered structures of the vitreous phase that are equivalent to the high-temperature non-crystalline structure in Fig. 3a.

Because it is extremely difficult to converge simulations involving assembly and diss-assembly processes, it is useful to compute the errors of the free energy surfaces by block averaging techniques.[79] The overall sampling time were 100 ns for all simulations. While the estimated block-averaged free energy surfaces of the thermal decomposition process are shown in Fig. 6a, the corresponding errors are reported in Fig. S10. The three projections of the free energy surface for each pair of CVs for a $2 \times 2 \times 2$ ZIF-8 supercell and 50 MeOH molecules per unit cell are shown in Fig. 6a. The ZIF-8 crystal corresponds to the region labeled as ($C$) (from crystalline). A second minimum can be observed, it is labeled as ($G$) (from glassy/vitreous), this potential well is characterized by very low values of the order-related CVs (from 0 to 0.15 for $\omega_{Q_6}^o$, and $\omega_{\text{sod}}^o \sim 0$), with relatively high values of $\kappa_1^c$. We can conclude that these two phases are distinct from the analysis of the error bars. Even though we cannot compute the free energy of the $C \leftrightarrow G$ transition due to the lack of sufficient sampling of the vitreous state, we can estimate the activation energy as the free energy barrier between the crystalline phase minimum ($G(C)$) and the transition state, $G^*$, that leads to the vitreous phase ($G$), $\Delta G_{C \to G}^{\dagger} = G^* - G(C)$. These activation energies are of 1.19, 1.58, and 1.55 respectively for $\kappa_1^c$, $\omega_{\text{sod}}^o$, and $\omega_{Q_6}^o$. We have also calculated the optimal pathway on the 2D-projections of the FES, minima, and transition state by using the algorithm MULE (stands for Multidimensional Low-est Energy) developed by Fu *et al.*[89] Typical snapshots of the system in the $(C)$ and $(G)$ phases are illustrated at the bottom of Fig. 6. Simulations containing less MeOH molecules (40 molecules per unit cell) exhibit a similar behavior, but the free energy barriers were higher and although the system explored the $(G)$ state, it was not able to return to the crystalline phase $(C)$. For an even lesser number of solvent molecules of 37.5 molecules per unit cell, the system was only able to explore the crystalline phase during a 100 ns-long simulation. These results highlight the crucial role of the solvent in the self-assembly process.

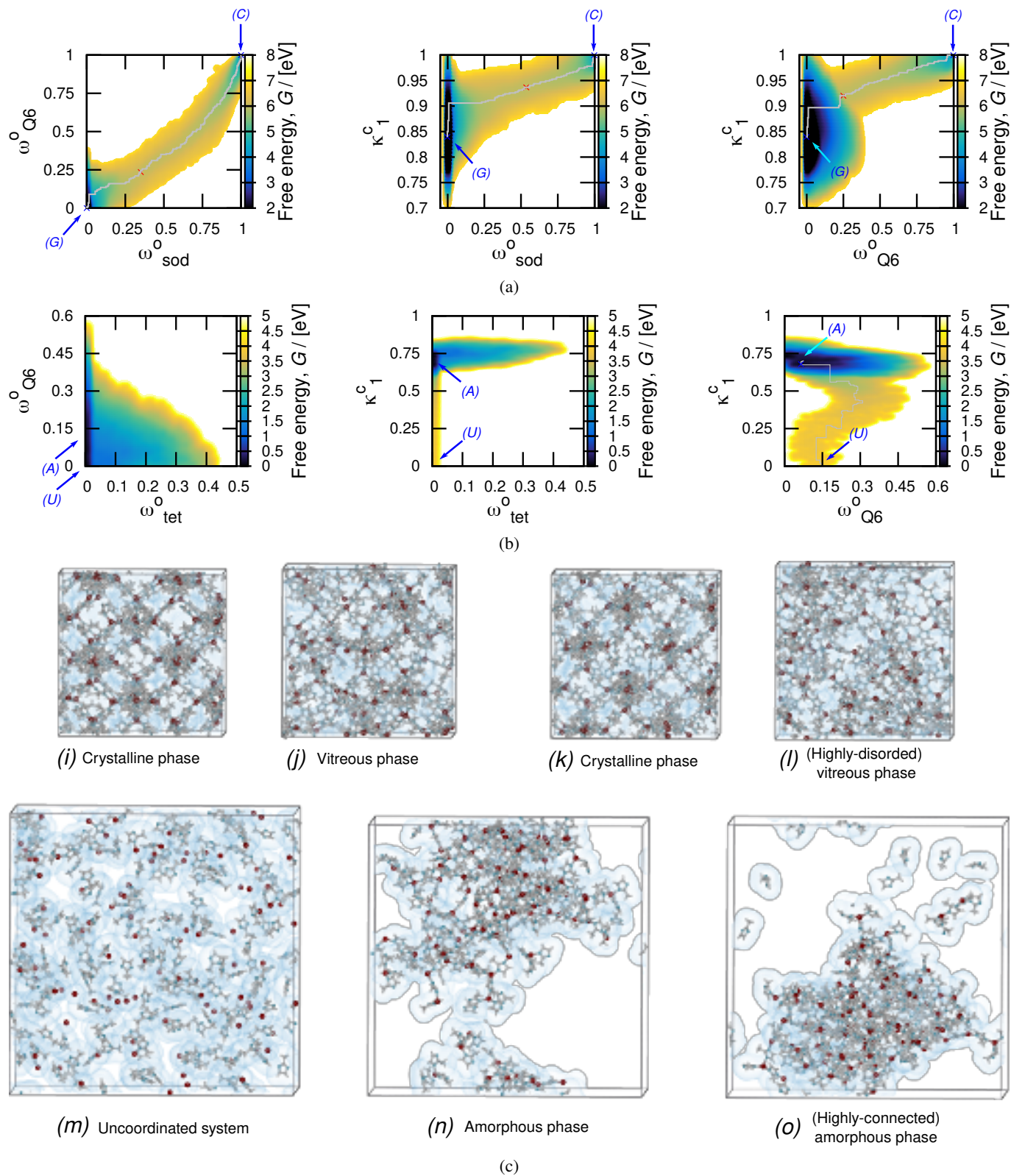

Figure 6: (a) Three projections of the FES for three CVs used in the WT-MetaD simulations ($\kappa_1^c$, $\omega_{Q_6}^o$, and $\omega_{sod}^o$) starting from the $2 \times 2 \times 2$ ZIF-8 crystal saturated with methanol at 450 K. Solid gray lines represent the optimal pathway on the 2D-FES, the blue points the minima (C) and (G) (crystalline and vitreous), and the red point the transition state (CG). (b) Three projections of the FES for the three CVs ($\kappa_1^c$, $\omega_{Q_6}^o$, and $\omega_{tet}^o$) from a simulation starting from the solvated $Zn^{2+}$ and $MIm^-$ moieties with 96 $Zn(MIm)_2$ 2240 MeOH composition at 450 K. The blue arrow to (A) marks the minima in the amorphous phase, (U) marks the uncoordinated initial microstate (not a phase), and the solid gray line represents the optimal pathway on the 2D-projection ($\kappa_1^c$, $\omega_{Q_6}^o$) of the FES. Note that this FES does not present a minima in the (U) point. (c) Snapshots of the trajectories for representative microstates for some of the phases visited along the simulations. Cell volumes are in the same scale. The (instantaneous) adsorption surface of water was added in translucent light blue. More supplementary snapshots are presented in Fig. S19.

Figs. S11 and S12 show a region in CV space that is visited during the WT-MetaD that has slightly lower values of connectivity than it is typical for the vitreous phase ($\kappa_2^c \sim 0.8$) associated to drastically reduced $\omega_{tet}^o$ values, as well as to an increase of the cell volume, $V$, (region labeled as highly-disordered vitreous phase in Fig. S12). While we cannot conclude that this region constitutes a separate phase from the vitreous one from the error analysis, there is an important qualitative difference between the associated configurations (see Fig. 6j and l).

Fig. 6b depicts the three projections of the free energy surface of the early polymerization process described in the previous section for comparison (error-bars are in Fig. S14 and Fig. S15). These simulations start from the uncoordinated $Zn^{2+}$ and $MIm^-$ solvated moieties (labeled $(U)$). The trajectories of all the walkers quickly converge to state $(A)$ (from *amorphous*), which is characterized by a high Zn-N connectivity but relatively low values of the order-related CVs $\omega_{tet}^o < 0.1$ and $\omega_{Q6}^o < 0.3$.

The connection between the amorphous aggregate state $(A)$ and the vitreous state $(G)$ cannot be made due to the very nature of the simulation performed (NPT MD). As the number of solvent molecules cannot be changed, the system cannot percolate through the PBCs and become *crystalline* (or the other way around). As mentioned before, to model this part of the free energy surface one must enable changes in the number of solvent molecules with constant chemical potential, as proposed by Karmakar, Piaggi, and Parrinello.[90]

## IV. CONCLUSIONS

In this work, we developed a computational methodology to unveil mechanistic details pertaining to the $Zn^{2+}$ - $MIm^-$ moieties polymerization (early stages of nucleation) and the thermal decomposition of a ZIF-8 crystal. We started by developing nb-ZIF-FF, an appropriate force field for this task, including (i) a physically sound model for taking into account the formation and breaking of metal-ligand connections *via* a Morse potential functional form, and (ii) cationic and anionic dummy atom models to mimic the spatial charge distribution of the metal cation in a ligands field. We have carefully validated our force field by checking the stability of different ZIFs and associated aggregates and by computing radial distribution functions, cell parameters and elastic constants. Since nb-ZIF-FF successfully captures structure and energetics of all tested ZIFs, it can be applied to model the self-assembly and phase transformation of other ZIF polymorphs, as well as other studies that require including metal-ligand reactivity at system sizes larger than what can be achieved with *ab initio* resolution. We hope this contribution triggers further work in these directions. As a second step, we detail the collective variables that we selected for our well-tempered metadynamics studies, we explain our rationale for choosing them and test their appropriateness and non-redundancy. We study the effect of temperature and of the size of the system in the simulation results. We explored the use of different solvents at the development part of our work and finally kept an explicit united atom model of methanol for the well-tempered metadynamics simulations. We discuss several tricks-of-the-trade related to modeling self-assembly processes.

Our well-tempered metadynamics simulations indicate that ZIF-8's self-assembly starts with a rapid increase of the Zn-N connectivity: up to 85% of the total possible bonds were formed within the first 150 ps. This stage is followed by the evaporation of small clusters into its constituents which are further combined with the largest clusters, while large clusters merge. Finally, an amorphous phase is formed, and the local order of the system gets stuck. Further progress in the self-assembly would require performing the simulation in a constant chemical potential ensemble, so that local solvent concentration fluctuations at the surfaces of the nuclei are possible and the system can percolate through the periodic boundaries. The pre-nucleation building units observed are single 4-fold and 3-fold connected $Zn^{2+}$ cations with lifetimes in the order of picoseconds as well as 4-,5- and 6-membered rings with lifetimes in the nanosecond realm. The free ligands act as "templating agents" for the formation of the sodalite cages. Finally, the free energy surfaces of crystal decomposition and of the early polymerization are explored, and their errors are calculated by block averaging. Thermal decomposition leads to the rapid formation of more disordered structures, and the stabilization of a vitreous phase $(G)$. The activation energy of the thermal decomposition associated to the different CVs is of $\sim 1.2$ and $1.5$ . The system also visits a region in phase space associated with a slightly lower connectivity and much lower tetrahedrality, which can be described as a highly-disordered vitreous phase. Going from the vitreous phase to the amorphous phase $(A)$ formed in the early polymerization simulations is not possible with the current methodology, but will be the object of further work.

Our work unveils molecular-level mechanistic details of the early stages of self-assembly and crystal decomposition of ZIF-8 for the first time. This contributes to augmenting our fundamental understanding of the assembly and disassembly processes, which is crucial in the rational design of new MOFs as well as for other materials. We also hope that this contribution will encourage further work in the simulation of other related reactive processes in solution.

## V. SUPPLEMENTARY MATERIAL

The Supplementary material contains details of the nb-ZIF-FF development, PLUMED codes to compute the different collective variables explored in this work and for the WT-MetaD calculations as well as additional principal component analysis and WT-MetaD results. More supporting materials are made available at https://github.com/salrodgom/ZIFWTMetaDnbZIFFF.

## ACKNOWLEDGMENTS

The authors thank G. Maurin, S. Hamad, and A. R. Ruiz–Salvador for valuable discussions at the initial stages of this work. We thank F. Abarca for his invaluable help in R coding. SRGB was supported by grants FJC2018-035697-I funded by MCIN/AEI/10.13039/501100011033 (Ministerio de Ciencia e Innovación; Agencia Estatal de Investigación) and POSTDOC_21_00069 funded by Consejería de Transformación Económica, Industria, Conocimiento y Universidades, Junta de Andalucía. This work was granted access to the HPC resources of CINES under the allocation A0090911989 made by GENCI. RS thanks the European Research Council for funding the ERC Starting Grant MAGNIFY (project 101042514) that will allow her to continue developing the computational modeling of MOFs self-assembly in solvothermal conditions.

## DATA AVAILABILITY STATEMENT

The data that support the findings of this study are available within the article, its supplementary material, and openly available at https://github.com/salrodgom/ZIFWTMetaDnbZIFFF.

## Author contribution

SRGB developed the force field, wrote the FORTRAN codes, set up the simulation framework (LAMMPS/PLUMED/FORTRAN codes) and performed all unbiased MD simulations. RS designed the study and performed the PC analyses. SRGB and RS performed the WT-MetaD simulations, discussed the results and wrote the manuscript.

## Conflicts of interest

The authors have no conflicts to disclose.

## REFERENCES

[1] X. Zhang, Z. Chen, X. Liu, S. L. Hanna, X. Wang, R. Taheri-Ledari, A. Maleki, P. Li, and O. K. Farha, "A historical overview of the activation and porosity of metal–organic frameworks," Chem. Soc. Rev. **49**, 7406–7427 (2020).

[2] P.-L. Wang, L.-H. Xie, E. A. Joseph, J.-R. Li, X.-O. Su, and H.-C. Zhou, "Metal–organic frameworks for food safety," Chem. Rev. **119**, 10638–10690 (2019).

[3] S. Kempahanumakkagari, K. Vellingiri, A. Deep, E. E. Kwon, N. Bolan, and K.-H. Kim, "Metal–organic framework composites as electrocatalysts for electrochemical sensing applications," Coord. Chem. Rev. **357**, 105–129 (2018).

[4] M. Giménez-Marqués, T. Hidalgo, C. Serre, and P. Horcajada, "Nanostructured metal–organic frameworks and their bio-related applications," Coord. Chem. Rev. **307**, 342–360 (2016), chemistry and Applications of Metal Organic Frameworks.

[5] S. Dhaka, R. Kumar, A. Deep, M. B. Kurade, S.-W. Ji, and B.-H. Jeon, "Metal–organic frameworks (mofs) for the removal of emerging contaminants from aquatic environments," Coord. Chem. Rev. **380**, 330–352 (2019).

[6] M. Ding, R. W. Flaig, H.-L. Jiang, and O. M. Yaghi, "Carbon capture and conversion using metal–organic frameworks and mof-based materials," Chem. Soc. Rev. **48**, 2783–2828 (2019).

[7] G. Maurin, C. Serre, A. Cooper, and G. Férey, "The new age of MOFs and of their porous-related solids," Chem. Soc. Rev. **46**, 3104–3107 (2017).

[8] M. Ding, R. W. Flaig, H.-L. Jiang, and O. M. Yaghi, "Carbon capture and conversion using metal–organic frameworks and MOF-based materials," Chem. Soc. Rev. **48**, 2783–2828 (2019).

[9] Y. Du, H. Su, T. Fei, B. Hu, J. Zhang, S. Li, S. Pang, and F. Nie, "Structure–property relationship in energetic cationic metal–organic frameworks: New insight for design of advanced energetic materials," Cryst. Growth Des. **18**, 5896–5903 (2018).

[10] G. E. Gomez, E. V. Brusau, J. Sacanell, G. J. A. A. S. Illia, and G. E. Narda, "Insight into the metal content-structure-property relationship in lanthanide metal-organic frameworks: Optical studies, magnetism, and catalytic performance," Eur. J. Inorg. Chem. **2018**, 2452–2460 (2018).

[11] W. Li, X. Xia, M. Cao, and S. Li, "Structure–property relationship of metal–organic frameworks for alcohol-based adsorption-driven heat pumps via high-throughput computational screening," J. Mater. Chem.A **7**, 7470–7479 (2019).

[12] S. Dissegna, P. Vervoorts, C. L. Hobday, T. Düren, D. Daisenberger, A. J. Smith, R. A. Fischer, and G. Kieslich, "Tuning the mechanical response of metal–organic frameworks by defect engineering," J. Am. Chem. Soc. **140**, 11581–11584 (2018).

[13] D. S. Sholl and R. P. Lively, "Defects in metal–organic frameworks: Challenge or opportunity?" J. Phys. Chem. Lett. **6**, 3437–3444 (2015).

[14] M. Taddei, "When defects turn into virtues: The curious case of zirconium-based metal-organic frameworks," Coord. Chem. Rev. **343**, 1–24 (2017).

[15] M. Safaei, M. M. Foroughi, N. Ebrahimpoor, S. Jahani, A. Omidi, and M. Khatami, "A review on metal-organic frameworks: Synthesis and applications," TrAC - Trends Anal. Chem. **118**, 401–425 (2019).

[16] V. Guillerm and D. Maspoch, "Geometry mismatch and reticular chemistry: Strategies to assemble metal–organic frameworks with non-default topologies," J. Am. Chem. Soc. **141**, 16517–16538 (2019).

[17] H. Jiang, J. Jia, A. Shkurenko, Z. Chen, K. Adil, Y. Belmabkhout, L. J. Weselinski, A. H. Assen, D.-X. Xue, M. O'Keeffe, and M. Eddaoudi, "Enriching the reticular chemistry repertoire: Merged nets approach for the rational design of intricate mixed-linker metal–organic framework platforms," J. Am. Chem. Soc. **140**, 8858–8867 (2018).

[18] T. Wang, E. Lin, Y.-L. Peng, Y. Chen, P. Cheng, and Z. Zhang, "Rational design and synthesis of ultramicroporous metal-organic frameworks for gas separation," Coord. Chem. Rev. **423**, 213485 (2020).

[19] R. Freund, S. Canossa, S. M. Cohen, W. Yan, H. Deng, V. Guillerm, M. Eddaoudi, D. G. Madden, D. Fairen-Jimenez, H. Lyu, L. K. Macreadie, Z. Ji, Y. Zhang, B. Wang, F. Haase, C. Wöll, O. Zaremba, J. Andreo, S. Wuttke, and C. S. Diercks, "25 years of reticular chemistry," Angew. Chem. Int. Ed. (2021), 10.1002/anie.202101644.

[20] H. Jiang, D. Alezi, and M. Eddaoudi, "A reticular chemistry guide for the design of periodic solids," Nat. Rev. Mater. (2021), 10.1038/s41578-021-00287-y.

[21] S. Surblé, F. Millange, C. Serre, G. Férey, and R. I. Walton, "An EXAFS study of the formation of a nanoporous metal–organic framework: evidence for the retention of secondary building units during synthesis," Chem. Commun. , 1518 (2006).

[22] M. Shoaee, M. Anderson, and M. Attfield, "Crystal growth of the nanoporous metal-organic framework HKUST-1 revealed by in situ atomic force microscopy," Angew. Chem. Int. Ed. **47**, 8525–8528 (2008).

[23] J. Cravillon, R. Nayuk, S. Springer, A. Feldhoff, K. Huber, and M. Wiebcke, "Controlling zeolitic imidazolate framework nano- and microcrystal formation: Insight into crystal growth by time-resolved in situ static light scattering," Chem. Mater. **23**, 2130–2141 (2011).

[24] M. P. Attfield and P. Cubillas, "Crystal growth of nanoporous metal organic frameworks," Dalton Trans. **41**, 3869–3878 (2012).

[25] M. G. Goesten, E. Stavitski, J. Juan-Alcañiz, A. Martiñez-Joaristi, A. V. Petukhov, F. Kapteijn, and J. Gascon, "Small-angle x-ray scattering documents the growth of metal-organic frameworks," Catal. Today **205**, 120–127 (2013).

[26] J. P. Patterson, P. Abellan, M. S. Denny, C. Park, N. D. Browning, S. M. Cohen, J. E. Evans, and N. C. Gianneschi, "Observing the growth of metal–organic frameworks by in situ liquid cell transmission electron microscopy," J. Am. Chem. Soc. **137**, 7322–7328 (2015).

[27] K. S. Walton, "Movies of a growth mechanism," Nature **523**, 535–536 (2015).

[28] S. Van Cleuvenbergen, Z. J. Smith, O. Deschaume, C. Bartic, S. Wachsmann-Hogiu, T. Verbiest, and M. A. van der Veen, "Morphology and structure of zif-8 during crystallisation measured by dynamic angle-resolved second harmonic scattering," Nat. Commun. **9**, 3418 (2018).

[29] D. Saliba, M. Ammar, M. Rammal, M. Al-Ghoul, and M. Hmadeh, "Crystal growth of zif-8, zif-67, and their mixed-metal derivatives," J. Am. Chem. Soc. **140**, 1812–1823 (2018).

[30] M. Filez, C. Caratelli, M. Rivera-Torrente, F. Muniz-Miranda, M. Hoek, M. Altelaar, A. J. Heck, V. V. Speybroeck, and B. M. Weckhuysen, "Elucidation of the pre-nucleation phase directing metal-organic framework formation," Cell Rep. Physical Science **2**, 100680 (2021).

[31] M. Yoneya, S. Tsuzuki, and M. Aoyagi, "Simulation of metal–organic framework self-assembly," Phys. Chem. Chem. Phys. **17**, 8649–8652 (2015).

[32] D. Biswal and P. G. Kusalik, "Probing molecular mechanisms of self-assembly in metal–organic frameworks," ACS Nano **11**, 258–268 (2016).

[33] D. Biswal and P. G. Kusalik, "Molecular simulations of self-assembly processes in metal-organic frameworks: Model dependence," J. Chem. Phys. **147**, 044702 (2017).

[34] Y.-P. Pang, "Novel zinc protein molecular dynamics simulations: Steps toward antiangiogenesis for cancer treatment," J. Mol. Model. **5**, 196–202 (1999).

[35] S. Jawahery, N. Rampal, S. M. Moosavi, M. Witman, and B. Smit, "Ab initio flexible force field for metal–organic frameworks using dummy model coordination bonds," J. Chem. Theory Comput. **15**, 3666–3677 (2019).

[36] Y. J. Colón, A. Z. Guo, L. W. Antony, K. Q. Hoffmann, and J. J. de Pablo, "Free energy of metal-organic framework self-assembly," J. Chem. Phys. **150**, 104502 (2019).

[37] L. Kollias, D. C. Cantu, M. A. Tubbs, R. Rousseau, V.-A. Glezakou, and M. Salvalaglio, "Molecular level understanding of the free energy landscape in early stages of metal–organic framework nucleation," J. Am. Chem. Soc. **141**, 6073–6081 (2019).

[38] A. Laio and M. Parrinello, "Escaping free-energy minima," Proc. Natl. Acad. Sci. U.S.A. **99**, 12562–12566 (2002).

[39] D. C. Cantu, B. P. McGrail, and V.-A. Glezakou, "Formation mechanism of the secondary building unit in a chromium terephthalate metal–organic framework," Chem. Mater. **26**, 6401–6409 (2014).

[40] L. Kollias, R. Rousseau, V.-A. Glezakou, and M. Salvalaglio, "Understanding metal–organic framework nucleation from a solution with evolving graphs," J. Am. Chem. Soc. **144**, 11099–11109 (2022).

[41] C. Healy, K. M. Patil, B. H. Wilson, L. Hermanspahn, N. C. Harvey-Reid, B. I. Howard, C. Kleinjan, J. Kolien, F. Payet, S. G. Telfer, P. E. Kruger, and T. D. Bennett, "The thermal stability of metal-organic frameworks," Coord. Chem. Rev. **419**, 213388 (2020).

[42] J. Yan, J. C. MacDonald, A. R. Maag, F.-X. Coudert, and S. C. Burdette, "MOF decomposition and introduction of repairable defects using a photodegradable strut," Eur. J. Chem. **25**, 8393–8400 (2019).

[43] A. Barducci, G. Bussi, and M. Parrinello, "Well-tempered metadynamics: A smoothly converging and tunable free-energy method," Phys. Rev. Lett. **100** (2008), 10.1103/physrevlett.100.020603.

[44] G. Bussi and D. Branduardi, "Free-energy calculations with metadynamics: Theory and practice," in *Rev. Comput. Chem.* (John Wiley & Sons, Inc, 2015) pp. 1–49.

[45] D. Quigley and P. Rodger, "A metadynamics-based approach to sampling crystallisation events," Mol. Simul. **35**, 613–623 (2009).

[46] Y.-Y. Zhang, H. Niu, G. Piccini, D. Mendels, and M. Parrinello, "Improving collective variables: The case of crystallization," J. Chem. Phys. **150**, 094509 (2019).

[47] T. Karmakar, M. Invernizzi, V. Rizzi, and M. Parrinello, "Collective variables for the study of crystallisation," Mol. Phys. **119** (2021), 10.1080/00268976.2021.1893848.

[48] K. S. Park, Z. Ni, A. P. Côté, J. Y. Choi, R. Huang, F. J. Uribe-Romo, H. K. Chae, M. O'Keeffe, and O. M. Yaghi, "Exceptional chemical and thermal stability of zeolitic imidazolate frameworks," Proc. Natl. Acad. Sci. U.S.A. **103**, 10186–10191 (2006).

[49] B. Chen, Z. Yang, Y. Zhu, and Y. Xia, "Zeolitic imidazolate framework materials: recent progress in synthesis and applications," J. Mater. Chem. A **2**, 16811–16831 (2014).

[50] D. Zacher, O. Shekhah, C. Wöll, and R. A. Fischer, "Thin films of metal–organic frameworks," Chem. Soc. Rev. **38**, 1418–1429 (2009).

[51] Y. Pan, Y. Liu, G. Zeng, L. Zhao, and Z. Lai, "Rapid synthesis of zeolitic imidazolate framework-8 (zif-8) nanocrystals in an aqueous system," Chem. Commun. **47**, 2071–2073 (2011).

[52] J. Cravillon, S. Münzer, S.-J. Lohmeier, A. Feldhoff, K. Huber, and M. Wiebcke, "Rapid room-temperature synthesis and characterization of nanocrystals of a prototypical zeolitic imidazolate framework," Chem. Mater. **21**, 1410–1412 (2009).

[53] X. Feng, T. Wu, and M. A. Carreon, "Synthesis of ZIF-67 and ZIF-8 crystals using DMSO (dimethyl sulfoxide) as solvent and kinetic transformation studies," J. Cryst. Growth **455**, 152–156 (2016).

[54] X.-G. Wang, Q. Cheng, Y. Yu, and X.-Z. Zhang, "Controlled nucleation and controlled growth for size predicable synthesis of nanoscale metal–organic frameworks (mofs): A general and scalable approach," Angew. Chem. Int. Ed. **57**, 7836–7840 (2018).

[55] S. R. Venna, J. B. Jasinski, and M. A. Carreon, "Structural evolution of zeolitic imidazolate framework-8," J. Am. Chem. Soc. **132**, 18030–18033 (2010).

[56] X. Liu, S. W. Chee, S. Raj, M. Sawczyk, P. Král, and U. Mirsaidov, "Three-step nucleation of metal–organic framework nanocrystals," Proc. Natl. Acad. Sci. U.S.A. **118**, e2008880118 (2021).

[57] S. Plimpton, "Fast parallel algorithms for short-range molecular dynamics," J. Comput. Phys. **117**, 1–19 (1995).

[58] E. Polak and G. Ribière, "Note sur la convergence de méthodes de directions conjuguées," Rev. Fr. Autom., Inf. Rech. Opér., Sér.: Rouge **3**, 35–43 (1969).

[59] E. Bitzek, P. Koskinen, F. Gähler, M. Moseler, and P. Gumbsch, "Structural relaxation made simple," Phys. Rev. Lett. **97** (2006), 10.1103/physrevlett.97.170201.

[60] D. Dubbeldam, S. Calero, D. E. Ellis, and R. Q. Snurr, "RASPA: molecular simulation software for adsorption and diffusion in flexible nanoporous materials," Mol. Simul. **42**, 81–101 (2015).

[61] P. Raiteri, A. Laio, F. L. Gervasio, C. Micheletti, and M. Parrinello, "Efficient reconstruction of complex free energy landscapes by multiple walkers metadynamics," J. Phys. Chem. B **110**, 3533–3539 (2006).

[62] Z. Hu, L. Zhang, and J. Jiang, "Development of a force field for zeolitic imidazolate framework-8 with structural flexibility," J. Chem. Phys. **136**, 244703 (2012).

[63] L. Zhang, Z. Hu, and J. Jiang, "Sorption-induced structural transition of zeolitic imidazolate framework-8: A hybrid molecular simulation study," J. Am. Chem. Soc. **135**, 3722–3728 (2013).

[64] B. Zheng, M. Sant, P. Demontis, and G. B. Suffritti, "Force field for molecular dynamics computations in flexible ZIF-8 framework," J. Phys. Chem. C **116**, 933–938 (2012).

[65] X. Wu, J. Huang, W. Cai, and M. Jaroniec, "Force field for ZIF-8 flexible frameworks: atomistic simulation of adsorption, diffusion of pure gases as CH4, h2, CO2 and n2," RSC Adv. **4**, 16503–16511 (2014).

[66] T. Weng and J. R. Schmidt, "Flexible and transferable ab initio force field for zeolitic imidazolate frameworks: ZIF-FF," J. Phys. Chem. A **123**, 3000–3012 (2019).

[67] J. P. Dürholt, G. Fraux, F.-X. Coudert, and R. Schmid, "Ab initio derived force fields for zeolitic imidazolate frameworks: MOF-FF for ZIFs," J. Chem. Theory Comput. **15**, 2420–2432 (2019).

[68] V. Nguyen and M. Grünwald, "Microscopic origins of poor crystallinity in the synthesis of covalent organic framework COF-5," J. Am. Chem. Soc. **140**, 3306–3311 (2018).

[69] B. L. Eggimann, A. J. Sunnarborg, H. D. Stern, A. P. Bliss, and J. I. Siepmann, "An online parameter and property database for the trappe force field," Mol. Simul. **40**, 101–105 (2014).

[70] D. W. Lewis, A. R. Ruiz-Salvador, A. Gómez, L. M. Rodriguez-Albelo, F.-X. Coudert, B. Slater, A. K. Cheetham, and C. Mellot-Draznieks, "Zeolitic imidazole frameworks: structural and energetics trends compared with their zeolite analogues," CrystEngComm **11**, 2272 (2009).

[71] E. C. Spencer, R. J. Angel, N. L. Ross, B. E. Hanson, and J. A. K. Howard, "Pressure-induced cooperative bond rearrangement in a zinc imidazolate framework: A high-pressure single-crystal x-ray diffraction study," J. Am. Chem. Soc. **131**, 4022–4026 (2009).

[72] J. R. Ramirez, H. Yang, C. M. Kane, A. N. Ley, and K. T. Holman, "Reproducible synthesis and high porosity of mer-zn(im)2 (ZIF-10): Exploitation of an apparent double-eight ring template," J. Am. Chem. Soc. **138**, 12017–12020 (2016).

[73] S. Cao, T. D. Bennett, D. A. Keen, A. L. Goodwin, and A. K. Cheetham, "Amorphization of the prototypical zeolitic imidazolate framework ZIF-8 by ball-milling," Chem. Commun. **48**, 7805 (2012).

[74] T. D. Bennett, S. Cao, J. C. Tan, D. A. Keen, E. G. Bithell, P. J. Beldon, T. Friscic, and A. K. Cheetham, "Facile mechanosynthesis of amorphous zeolitic imidazolate frameworks," J. Am. Chem. Soc. **133**, 14546–14549 (2011).

[75] J.-C. Tan, B. Civalleri, C.-C. Lin, L. Valenzano, R. Galvelis, P.-F. Chen, T. D. Bennett, C. Mellot-Draznieks, C. M. Zicovich-Wilson, and A. K. Cheetham, "Exceptionally low shear modulus in a prototypical imidazole-based metal-organic framework," Phys. Rev. Lett. **108** (2012), 10.1103/physrevlett.108.095502.

[76] B. Zheng, Y. Zhu, F. Fu, L. L. Wang, J. Wang, and H. Du, "Theoretical prediction of the mechanical properties of zeolitic imidazolate frameworks (ZIFs)," RSC Adv. **7**, 41499–41503 (2017).

[77] J. Cravillon, C. A. Schröder, H. Bux, A. Rothkirch, J. Caro, and M. Wiebcke, "Formate modulated solvothermal synthesis of zif-8 investigated using time-resolved in situ x-ray diffraction and scanning electron microscopy," CrystEngComm **14**, 492–498 (2012).

[78] The PLUMED consortium, "Promoting transparency and reproducibility in enhanced molecular simulations," Nat. Methods **16**, 670–673 (2019).

[79] G. Bussi and G. A. Tribello, "Analyzing and biasing simulations with PLUMED," in *Methods Mol. Biol.* (Springer New York, 2019) pp. 529–578.

[80] G. A. Tribello, F. Giberti, G. C. Sosso, M. Salvalaglio, and M. Parrinello, "Analyzing and driving cluster formation in atomistic simulations," J. Chem. Theory Comput. **13**, 1317–1327 (2017).

[81] P. M. Piaggi and M. Parrinello, "Calculation of phase diagrams in the multithermal-multibaric ensemble," J. Chem. Phys. **150**, 244119 (2019).

[82] A. P. Bartók, R. Kondor, and G. Csányi, "On representing chemical environments," Phys. Rev. B **87** (2013), 10.1103/physrevb.87.184115.

[83] S. De, A. P. Bartók, G. Csányi, and M. Ceriotti, "Comparing molecules and solids across structural and alchemical space," Phys. Chem. Chem. Phys. **18**, 13754–13769 (2016).

[84] P. J. Steinhardt, D. R. Nelson, and M. Ronchetti, "Bond-orientational order in liquids and glasses," Phys. Rev. B **28**, 784–805 (1983).

[85] P. R. ten Wolde, M. J. Ruiz-Montero, and D. Frenkel, "Numerical calculation of the rate of crystal nucleation in a lennard-jones system at moderate undercooling," J. Chem. Phys. **104**, 9932–9947 (1996).

[86] W. Lechner and C. Dellago, "Accurate determination of crystal structures based on averaged local bond order parameters," J. Chem. Phys. **129**, 114707 (2008).

[87] D. Dubbeldam, S. Calero, and T. J. Vlugt, "iRASPA: GPU-accelerated visualization software for materials scientists," Mol. Simul. **44**, 653–676 (2018).

[88] R. Murugavel, M. G. Walawalkar, M. Dan, H. W. Roesky, and C. N. R. Rao, "Transformations of molecules and secondary building units to materials: a bottom-up approach," Acc. Chem. Res. **37**, 763–774 (2004).

[89] H. Fu, H. Chen, X. Wang, H. Chai, X. Shao, W. Cai, and C. Chipot, "Finding an optimal pathway on a multidimensional free-energy landscape," J. Chem. Inf. Model. **60**, 5366–5374 (2020).

[90] T. Karmakar, P. M. Piaggi, and M. Parrinello, "Molecular dynamics simulations of crystal nucleation from solution at constant chemical potential," J. Chem. Theory Comput. **15**, 6923–6930 (2019).